\documentclass[structabstract]{aa}  
\usepackage{graphicx,amssymb,amsmath}
\usepackage{natbib}
\usepackage{txfonts}

\def\arcdeg{$^{\circ}$}
\def\H0{{\rm ~km~s^{-1}~Mpc^{-1}}}

\def\850{850$\mu$m}

\def\hii{\ion{H}{II}}

\def\.25{0.25 keV\thinspace}

\def\z2{z$\,\sim\,2$}

\newcommand\dummytable{\refstepcounter{table}}%

\begin{document}
   \title{Ultra-High Energy Cosmic Rays Detected by Auger and AGASA:}
   \subtitle{Corrections for Galactic Magnetic Field Deflections, Source Populations, and 
             Arguments for Multiple-Components}
   \author{Neil M. Nagar
          \inst{1}
          \and
          Javier Matulich
          \inst{1}
          }
   \offprints{Neil M. Nagar}
   \institute{Astronomy Department, Universidad de Concepci\'on, Concepci\'on, Chile \\ 
              \email{nagar@astro-udec.cl,jmatulich@udec.cl}
             }
   \date{Received June 16, 2009; accepted December 5, 2009}

  \abstract
  % context heading (optional)
   {The origin and composition of Ultra-High Energy Cosmic Ray Events (UHECRs)
    are under debate. Possible sources include Active Galactic Nuclei - selected
    by various criteria - and extragalactic magnetars.
   }
  % aims heading (mandatory)
   {We aim to improve constraints on the source population(s) and
    compositions of UHECRs by accounting for UHECR deflections within 
    existing Galactic magnetic field models (GMFs).
   }
  % methods heading (mandatory)
   {We used Monte Carlo simulations for UHECRs detected by the Pierre Auger Observatory 
    and AGASA, to determine the UHECR trajectories
    within the galaxy and their outside-the-Galaxy arrival directions.
    The simulations, which used UHECR compositions from protons
    to Iron and seven models of the ordered GMF, 
    accounted for uncertainties in the GMF and a turbulent magnetic field component.
    The trajectories and outside-the-Galaxy arrival directions were compared
    with Galactic and extragalactic sources.
   }
  % results heading (mandatory)
   {For a given proton or light UHECR, the multiple potential outside-the-Galaxy arrival
    directions within a given GMF model are not very different, allowing meaningful constraints
    on source populations. 
    The correlation between a subset of UHECRs and nearby extended
    radiogalaxies (Nagar \& Matulich 2008) remains valid, even strengthened,
    within several GMF models. Both the nearest 
    radiogalaxy Cen~A, and the nearest radio-extended BL~Lac,
    CGCG~413$-$019, are potentially sources of multiple UHECRs.
    The correlation appears to be linked to the presence of the extended
    radio source rather than a tracer of an underlying matter distribution.
    Several UHECRs have trajectories which pass close to or through the 
    Galactic plane, some passing close to Galactic magnetars and/or 
    microquasars. For heavier UHECRs, the multiple potential outside-the-Galaxy arrival 
    directions of any given UHECR are highly scattered but still allow meaningful
    constraints. It is possible,
    but unlikely, that all UHECRs originate in the nearby radiogalaxy Cen~A.  
   }
  % conclusions heading (optional), leave it empty if necessary 
   {Nearby radiogalaxies remain a strong potential source
    of a significant subset of UHECRs.
    For light UHECRs about a third of UHECRs can be ``matched'' to nearby galaxies 
    with extended radio jets. The remaining UHECRs could also be explained 
    as originating in extended radiogalaxies if one has at least one of:
    a large UHECR mean free path, 
    a high cluster and/or intergalactic magnetic field,
    a heavy composition for two-thirds of the detected UHECRs. 
    If extended radiogalaxies are, or trace, UHECR sources, the most 
    consistent models for the ordered GMF are the BS-S and BS-A models; 
    the GMF models of Sun et al. 2008 are acceptable if a dipole component is added.
    }

   \keywords{interstellar-medium: Cosmic Rays -- interstellar-medium: Magnetic Fields --
             Galaxies: Active -- Galaxies: jets;}
   \authorrunning{Nagar \& Matulich}
   \titlerunning{UHECRs: Galactic Magnetic Field Deflections}
   \maketitle

\section{Introduction}
\label{secintro}

Ultra-High Energy Cosmic Rays (UHECRs) are protons or fully ionized
nuclei with energies greater than about 10$^{19}$ eV (10 EeV). 
On entering the earth's atmosphere they produce a shower of secondary particles 
and excite atmospheric molecules. The detection of both effects from
the ground allows a precise measurement of both the initial energy and the
arrival direction of the UHECR above the earth's atmosphere 
e.g. AGASA \citep{taket99} and 
the Pierre Auger Observatory \citep{abret04}. 
The resultant positional accuracy of the earth-arrival direction, 
now better than a degree, 
allows correlations with astronomical sources in order to
determine the source population(s) of UHECRs - a long held
mystery. Exploiting the full potential of the earth-arrival directions
nevertheless requires taking into account deflections suffered 
due to Galactic and extragalactic magnetic fields.

Suspected sources of ultra high energy cosmic rays include 
powerful active galaxies, magnetars,
core collapse Supernovae (SN II), and
gamma rays bursts (for reviews and lectures see
\citet{hil99, kat08} and references therein).
Specifically, an origin
in nearby radiogalaxies and BL~Lacs has long been predicted
\citep{racbie93, romet96}.
More exotic explanations include dark matter anhilation
and evaporation of micro black holes. 
The main energy loss for a UHECR propagating over cosmological 
distances is expected to be pion-production triggered by
interaction with a CMB photon, the so called 
Greisen Zatsepin Kuzmin  (GZK) effect \citep{gre66, zatkuz66}.
The predicted energy loss (around 30\%) and 
mean free path of the UHECR ($\sim$20--100~Mpc) depend on the 
composition and energy of the UHECR and details of the collision
process, e.g.  the photo-proton cross section 
(see e.g. \citet{sta04}; \citet{kat08}; PA08). 

The Pierre Auger Observatory (PAO) has detected 81 events between 
2004 January 1 and 2007 August 31  with reconstructed energies above 
40~EeV and zenith angles smaller than 60\arcdeg; of these, 27 have energies
above 56~EeV and have been reported in 
\citet[][hereafter PA07, PA08]{abret07, abret08}. 
The origins of the latter 27 UHECRs, together with 11 UHECRs detected by 
AGASA at energies above 56~EeV \citep{hayet00} are the focus of this paper.
The PAO has since detected additional UHECRs \citep[][hereafter PA09a, PA09b]{abret09a, abret09b}
but detailed positions and energies are not yet published.
In this work we use ``UHECR'' to refer to UHECRs with energy
above 56~EeV unless explicitly mentioned otherwise. 
The compositions of the detected UHECRs are still under debate
\citep[e.g. PA09a; PA09b; ][]{mat07, hootay09} with evidence for a 
mix of protons and heavy nuclei at the highest energies. 
There has been recent doubt about the measured
energies of the UHECRs detected by PAO and AGASA. The PAO team
have lowered their previous reported energies by a small percentage
(PA09a, PA09b) 
and there have been suggestions that energies reported by AGASA 
should be lowered by up to 30\%. To avoid confusion we report our
results separately for PAO and AGASA UHECRs.
Several other UHECR observatories, e.g. Haverah Park and Fly's Eye, have
detected a significant number of UHECRs: given that the GMF deflection 
calculations and source matching in this work require accurate energies 
($\lesssim$10\%) and arrival directions ($\lesssim$2\arcdeg) we 
do not include their results in our analysis. 

The arrival directions of the 27 UHECRs detected by PAO are
not isotropic at a 99\% significance level (PA07, PA08, PA09a, PA09b).
Several correlations, or lack thereof, with extragalactic sources have been suggested:
AGNs from the catalog of V{\'e}ron-Cetty \& V{\'e}ron (PA07, PA08),
hard X-ray selected AGNs \citep{geoet08}, nearby extended radiogalaxies
\citep{nagmat08}, nearby spiral galaxies \citep{ghiet08}, 
local volume galaxies \citep{cuepra09} and large scale
structure \citep{ryuet09}.

Galactic and extragalactic magnetic fields deflect cosmic rays
via the Lorentz force even at energies above 56~EeV. If the
extragalactic field is unordered over large scales the expected
deflections for these energies are only a few degrees \citep[e.g.][]{haret02,dolet04}. 
An ordered Galactic Magnetic Field (GMF) can, however, produce significant
deflections \citep[e.g.][]{taksat08} 
especially on trajectories which pass close to the Galactic plane
and/or for heavy composition UHECRs. 

The line of sight integrated GMF of the Galaxy has been
measured in various directions via e.g. the rotation measures 
of pulsars and extragalactic radio sources, and polarization
of starlight or radio synchrotron emission 
\citep[e.g. ][]{hei96, broet07, nouet08, sunet08, han09, bec09}. Additionally,
local magnetic fields can be measured accurately via Zeeman splitting
of emission lines.
For a given line of sight, the rotation measure traces the
integrated parallel component of the GMF 
while a UHECR is deflected by the perpendicular component of the 
GMF. A global model of the ordered GMF is thus required to transfer information 
from rotation measures into information useful for UHECR deflections. 
Several models have been presented for ordered magnetic fields
in the Galaxy. These typically consider one or more of the
following: a toroidal type field in and near
the galaxy disk, a toroidal field in the halo, and a poloidal field
which reproduces the vertical component seen in the Solar 
neighbourhood and the Galactic center. Models for the ordered
component of the magnetic field are discussed in the next section,
and recent comprehensive reviews can be found in \citet{han08}
and \citet{bec08}.
Studies of the global magnetic field of other galaxies has greatly 
facilitated the refinement of models of the Galactic GMF
\citep[see][for a review]{bec08, bec09}, even though there is significant
variation between galaxies in both the ordered and turbulent GMF, 
especially among those with high star formation rates.

In this article, we expand the work of \citet{nagmat08} in
two directions: first we use Monte Carlo simulations to
derive UHECR trajectories within the Galaxy and corresponding
outside-the-Galaxy arrival directions for all PAO and AGASA UHECRs.
The simulations, run for six UHECR compositions - from protons 
to iron - and seven GMF models, include uncertainties
in the GMF and a turbulent magnetic field component.
We then compare the UHECR trajectories within the
galaxy and their outside-the-Galaxy arrival directions with various
Galactic and extragalactic sources, and discuss the
implications of the results. 
The individual steps followed in this work have been addressed
previously by various authors referenced here and in the
next section. The new facet
of this work is that we combine all of the above
steps within the same Monte Carlo simulation, in order to model 
the trajectories of the PAO and AGASA UHECRs with energies greater
than 56~EeV within a large number of  ordered GMF models and
UHECR compositions.  
Sec.~\ref{secmonte} introduces the magnetic field models 
and describes the Monte Carlo simulations, 
Sec.~\ref{secdata} summarizes the sources of the data used,
and 
Sec.~\ref{secres} describes the principle results obtained. 
Finally, Sec.~\ref{secdis} contains a 
brief discussion and the conclusions of our study.
Distances to galaxies are calculated using a Hubble constant
of 72 $\H0$, except for 
relatively nearby galaxies for which we use
distances as referenced.

\section{Magnetic Fields and Monte Carlo Simulations}
\label{secmonte}

We consider seven ordered magnetic field
models for our Galaxy \citep[for recent reviews of Galactic
and extragalactic ordered magnetic fields see ][]{bec08, han08, bec09}.
Four of these models, the so called
BS-S, BS-A, AS-S, and AS-A have been used previously
by several authors.  Here, the first two alphabets
signify a (bi-)symmetric (BS) or asymmetric (AS) configuration w.r.t. 
a transformation between $\theta$ and $\theta+\pi$, where $\theta$ is the 
cylindrical azimuthal angle in the Galactic plane.
The last alphabet signifies whether
the field reverses direction (A) or not (B) in passing from above
the Galactic plane to below the Galactic plane.
Given the evidence for a vertical ($z$) component of the GMF, both in the Galactic
center and in the Solar vicinity, a dipole field model is 
usually added to the toroidal field above. Detailed descriptions 
and discussions of these GMFs can be found in 
\citet{sta97, bec01, han02, tintka02, prosmi03, broet07, kacet07, menet08, taksat08, han09}.
For these four models, we use the equations and their related 
normalizations strictly following \citet{taksat08}, and always
include the dipole field component as specified in \citet{taksat08}. 

\citet[][hereafter S08]{sunet08} have
recently presented three new variations of the above GMF 
models - AS-S+RING, AS-S+ARM and a modified BS-S - 
tailored to best fit observations of Galactic synchrotron emission
and its polarization, and rotation measures of pulsars. 
For brevity we sometimes refer to these three models as ASS+R, ASS+A, and BSS(S)
in the text and tables.
\citet{sunet08} find that the first
two models provide the best fit to current radio observations.
These two models use the basic configuration of the 
AS-S model, but introduce field reversals in a specified ring or
arm, respectively. The modified BS-S model is based on the 
traditional BS-S model but with 
different parameters and normalizations. The three new models
presented by S08 also include a toroidal halo field 
as initially presented in \citet{prosmi03}, but do not include
a dipole magnetic field component.
We use these three models and their related halo field as described in 
S08 with only one difference: in the AS-S+ARM model we directly 
use the log spiral arm morphology described in \citet{waiet92} without 
correcting for their shape near the solar neighbourhood as in 
S08, who used the correction suggested in \citet{taycor93}.
We do not expect significant changes in the UHECR trajectories due to 
this small change in the morphology of the spiral arms. 
\citet{sunet08} assumed that there is no ordered dipole field component
in the Galaxy GMF, arguing that the vertical ($z$) component of the local magnetic field
is a turbulent component. Since the parametric form and normalizations of the 
S08 models were derived via fits to radio observations, it is not strictly correct to add 
a dipole field to the models. Nevertheless, for illustrative purposes (and out
of curiosity) we also performed simulations of the S08 GMF models 
including the dipole field component as specified in \citet{taksat08}.
We refer to these results as the Sun et al. (2008) GMFs plus dipole.

We simulated UHECR trajectories for six different UHECR compositions: 
protons,
He$^{2+}$, 
O$^{8+}$, 
Al$^{13+}$, 
Ca$^{20+}$, and
Fe$^{26+}$. 
Once chosen, a composition was maintained fixed during the full
trajectory within the galaxy: i.e. we do not consider 
photo-dissociation-produced composition changes to a UHECR during its Galactic 
trajectory.
UHECR trajectories were simulated by firing an anti-particle with
the corresponding UHECR mass, (anti)charge, and energy from the Earth in the 
earth-arrival direction of the UHECR and following its
trajectory - including deflections in the magnetic field - in steps of 
10~pc. 
The simulations were carried out until the anti-particle
left a sphere with diameter 40~kpc centered on the Galactic center.
The Galactic coordinate pair ($l,b$) derived from the final position
and velocity of the anti-particle in the simulation is equivalent
to the arrival direction of the UHECR before it entered the Galaxy,
and we refer to this direction as the `outside-the-Galaxy' arrival direction,
to distinguish it from the `earth-arrival' direction measured by the
observatory.

Each UHECR's trajectory within each GMF model 
was simulated fifty times using a Monte Carlo 
approach: introducing uncertainties in the ordered magnetic field and
a random turbulent magnetic field component. 
In more detail, in each 10~pc simulation step we 
(a)~computed the 3-vector representing the Cartesian coordinates of the ordered
    magnetic field from the relevant model. 
    For the AS-S, AS-A, BS-S, and BS-A models this was the sum of the cylindrical 
    (disk) field and the dipole field. 
    For the AS-S+ARM, AS-S+RING, and BS-S model of S08 
    this was the sum of the disk and halo fields (with or without the dipole
    field as described above); 
(b)~added a Gaussian-distributed random error in each 
    component of the ordered  magnetic field, i.e. a change in both magnitude and 
    direction of the ordered field. This was done by adding to
    the 3-vector of the previous step a 3-vector whose each component 
    was a random number derived from a Gaussian distribution with 
    mean~=~0 and $\sigma$~=~0.5 times the value of the corresponding Cartesian 
    component of the GMF;
(c)~added a turbulent field component.
    The turbulent field component was computed following the recipe described
    in \citet{prosmi03}, though with slightly different normalization parameters: 
    the maximum turbulent magnetic field strength in each
    Cartesian component was taken to be
    $\pm$10$\mu$G, and the probability of a turbulent field being present was taken to
    be 1\% in the halo of the galaxy, 20\% in the disk of the galaxy (Galactic
    radius less then 20~kpc and Galactic height less than 1.5~kpc) and 80\% in
    the spiral arms of the Galaxy - where we used the spiral arm definition from
    \citet{waiet92}, spiral arm thickness 0.75~kpc in the disk (i.e. coordinate
    $r$), and half-height 0.5~kpc (coordinate $z$).
    We therefore generated a random number to determine whether a turbulent field
    was required to be added following the probabilities above. If so,
    we generated a 3-vector in which each component was a random
    number uniformly distributed between $\pm10\mu$G. This was added to the
    magnetic field 3-vector in the previous step in order to derive the total
    magnetic field used in the deflection calculation of this step. This turbulent field was
    then kept constant for the next 50~pc (i.e. an additional five simulation steps), 
    the expected scale length of turbulent field in the Galaxy \citep{prosmi03, bec08}.
    Note that the turbulent field used here is relatively large in 
    comparison with current measurements of the turbulent
    field in the Galaxy (S08) and other galaxies \citep{bec09}. 
   
In summary, within each GMF model we derived 50 outside-the-Galaxy  arrival directions
for each UHECR of a specific composition, or a total of 300 outside-the-Galaxy
arrival directions for each UHECR over all six compositions modelled.

\section{Data}
\label{secdata}

Positions and energies of the 27 UHECRs detected by the PAO
with energies $\geq$~56~EeV were taken from PA08; we have not made
the small energy corrections to these UHECRs as later reported by the PAO
(PAO09b). 
Positions and energies of the UHECRs detected by AGASA with 
energies $\geq$~56~EeV were taken from \citet{hayet00} which is an updation of 
the data listed in \citet{taket99}. For AGASA UHECRs, we lowered the
energies reported in \citet{taket99} by 30\%: this resulted in 11 UHECRs with 
energy above 56~EeV. In this work we use `AGASA UHECRs' to refer to only these
11 UHECRs unless explicitly stated otherwise. 

Our catalogs of astronomical sources were drawn from various
publications and web-based source lists, and chosen to be
representative of previously suspected or posited Galactic
and extragalactic sources of UHECRs. In all cases, we attempted to select the
most comprehensive catalog available.
Source lists for galaxies, galaxy clusters, GRBs, Galactic supernova remnants,
and extragalactic radio supernova are described in \citet{nagmat08}.
As in \citet{nagmat08}, in the case of radiogalaxies and radio sources we used
several surveys and catalogs, including NVSS \citep{conet98},
SUMMS \citep{bocet99}, and NED. The list of extragalactic jets is based
on that compiled by \citet{liuzha02}; to this list we added data on the galaxies
as given in NED, and re-measured the total flux and total extent 
of the radio emission from NVSS and SUMMS maps or from the 
list of DRAGNs compiled by 
P. Leahy \footnote{http://www.jb.man.ac.uk/atlas/dragns.html}. The DRAGNs
list is a subset of the 3CRR sources \citep{laiet83} which
have one or all of radio-emitting jets, lobes, and hotspots.
The total extent of the radio emission therefore includes both
radio jets and any radio lobes. This extent is typically referred to as
the Largest Angular Size (LAS) or Largest Linear Size (LLS). 
In the case of double sided jet or lobe sources, 
we added the LLS of the two jets and lobes as scalars instead of vectors
in order to discount the effects of jet and lobe bending.
\citet{nagmat08} listed the 10 radiogalaxies in the field of view of PAO
with D$\leq$ 75~Mpc and LLS $\geq$ 180~kpc. To this we added all other (northern)
radiogalaxies satisfying the same criteria: 
NGC  315 (D=~69.5~Mpc; LLS=~1111~kpc),
NGC  383 (D=~71.7~Mpc; LLS=~907~kpc),
NGC 1275 (D=~74.1~Mpc; LLS=~468~kpc),
NGC 5127 (D=~68.3~Mpc; LLS=~230~kpc), and
CGCG 514-050 (D=~72~Mpc; LLS=~506~kpc). 
Together, these 15 galaxies, form our sample of all
nearby (D$\leq$ 75~Mpc) extended (LLS $\geq$ 180~kpc) radiogalaxies.
In the case of nearby galaxies, we also used the HIPASS catalog \citep{meyet04, wonet06}
as used by \citet{ghiet08} in their study of the correlation between
UHECR arrival directions and nearby galaxies. Additionally, we used the 
revised Third Reference Catalog of Bright Galaxies \citep{coret94} to 
select a sample of nearby elliptical galaxies.

In this work we have considered several new source catalogs: 
Galactic Soft Gamma-Ray Repeaters (SGRs) and Galactic Anamolous X-ray Pulsars
(AXPs) were taken from \citet{wootho06} and updated with 
the online list maintained by the McGill pulsar 
group\footnote{http://www.physics.mcgill.ca/~pulsar/magnetar/main.html}. As
of June 2009, this list includes six SGRs (two of which are candidates) and 
10 AXPs (one of which is a candidate). 
These SGRs and AXPs are believed to be high magnetic field radio pulsars 
or ``magnetars''. Of the total of 16 SGRs and AXPs, 14 are close to the plane 
of the Galaxy, and two are in the SMC and LMC.

Confirmed Galactic microquasars were taken from the list in \citet{par05},
and candidate micro-quasars from \citet{comet08}.
Various gamma-ray catalogs have been used:
the BeppoSAX catalog of GRB and X-ray afterglows \citep{depet06},
the BeppoSAX complete catalogue of GRBs \citep{vetet07},
Gamma-ray Blazars in northern sky \citep{sowet03},
Blazar counterparts for 3EG sources \citep{sowet04},
Gamma-ray Blazar candidates \citep{sowet05},
Sources detected by ISGRI \citep{bodet07},
and the Third EGRET catalog \citep{haret99}.
We also used the HESS \citep{hof05} catalog of 54 Gamma-ray sources between 
100~GeV and 100~TeV as obtained from the HESS online
catalog in June 2009\footnote{http://www.mpi-hd.mpg.de/hfm/HESS/pages/home/sources/}.

\section{Results}
\label{secres}

We performed several Monte Carlo test runs varying the magnitudes of the 
error of the ordered field and the maximum turbulent field strength.
Here we present and discuss the results of simulations which used
a Gaussian-distributed (mean~=~0 and $\sigma$=~50\% of the value of the Cartesian GMF 
component) error in each Cartesian component of the ordered GMF,
and a turbulent magnetic field uniformly distributed between  
$\pm10\mu$G in each Cartesian component of the magnetic field,
with details as described in Sec.~\ref{secmonte}. 
We note that for the simulated GMF error, using 25\%--100\% instead of 50\% 
as described above 
does not significantly change the trajectories in the case of anti-protons.
The relative deflections suffered by the UHECR from the ordered field and turbulent 
field components can thus be roughly judged by comparing the average and r.m.s. of
the deflection angles corresponding to 
the fifty outside-the-Galaxy arrival directions for each UHECR within each 
GMF model. 

We used the following distance limits between 
a single outside-the-Galaxy arrival direction
and an extragalactic astronomical object in order to be considered a ``match'':
2.5\arcdeg\ for H, 
3\arcdeg\ for He, 
6\arcdeg\ for O, 
9\arcdeg\ for Al, 
12\arcdeg\ for Ca, and
15\arcdeg\ for Fe. These semi-arbitrary values were chosen considering a constant additive 
error of 2\arcdeg\ - to account for e.g. the error in determining the arrival direction
of the UHECR - plus a composition-weighted error of 0.5 times the charge of the UHECR 
which roughly accounts for errors in the deflection calculations and deflections by
intergalactic magnetic fields but avoids too large ``match radii'' for heavy nuclei. 
To match an astronomical source with a UHECR we require that at least 4 of the
50 outside-the-Galaxy arrival directions are `matched' by the above criterion; this $\geq$8\% 
match probability was chosen as it implies roughly that a match cannot be ruled out at better 
than a 2$\sigma$ level within the simulation parameters. 

\begin{figure*}[!htp]
\resizebox{4.8in}{!}{
\includegraphics[width=3.5in,angle=90,clip]{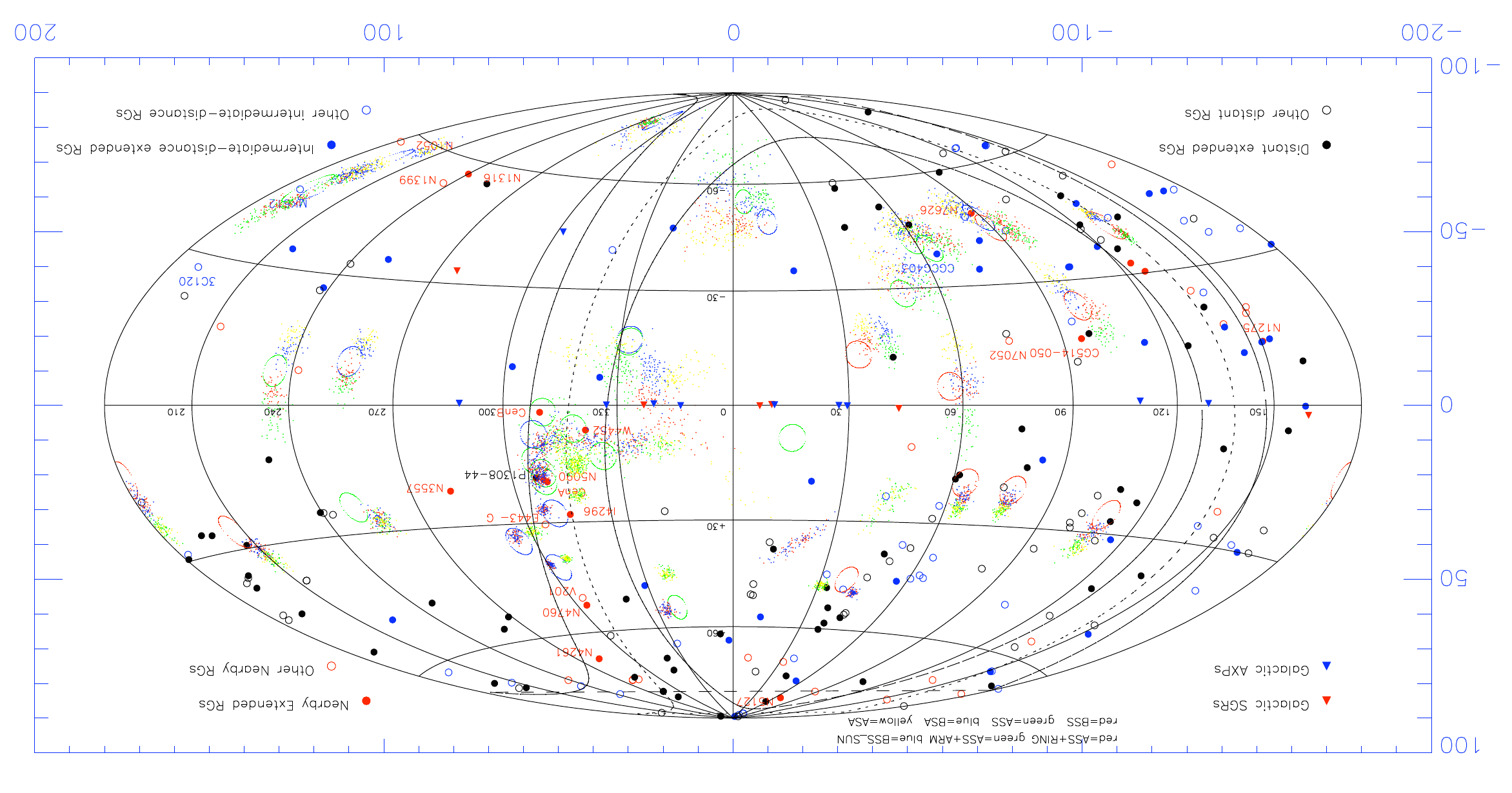}
}
\caption{A comparison of the earth-arrival directions of all UHECRs with
energies above 56~EeV detected by 
PAO (blue for UHECRs with energy $\geq$ 75~EeV and green for UHECRs with energy
between 56 and 75~EeV ) and AGASA (red open circles of radius 3.5\arcdeg)
with the estimated outside-the-Galaxy arrival directions of the same
UHECRs (small colored dots) for our Monte Carlo simulations
of the BS-S (red), AS-S (green), BS-A (blue), and AS-A (yellow) GMFs using
a proton composition for the UHECRs.
The colored circular symbols mark the positions of galaxies with radio jets at
D $\leq$ 75~Mpc (red),
75~Mpc $<$ D $\leq$ 200~Mpc (blue), and
200~Mpc $<$ D $\leq$ 500~Mpc (black).
In all redshift bins, filled circular symbols are used for galaxies with 
radio structures more extended
than 180~kpc, and open circular symbols for galaxies with radio structures less
extended than 180~kpc. 
Galactic SGRs (red triangles) and AXPs (blue triangles) are also plotted.
The Supergalactic plane is marked by the dashed line. In this and following
figures we use an Aitoff-Hammer (equi-area) projection in Galactic coordinates.
}
\label{figjet0123}
\end{figure*}

\begin{figure*}[!ht]
\resizebox{4.8in}{!}{
\includegraphics[width=3.5in,angle=90,clip]{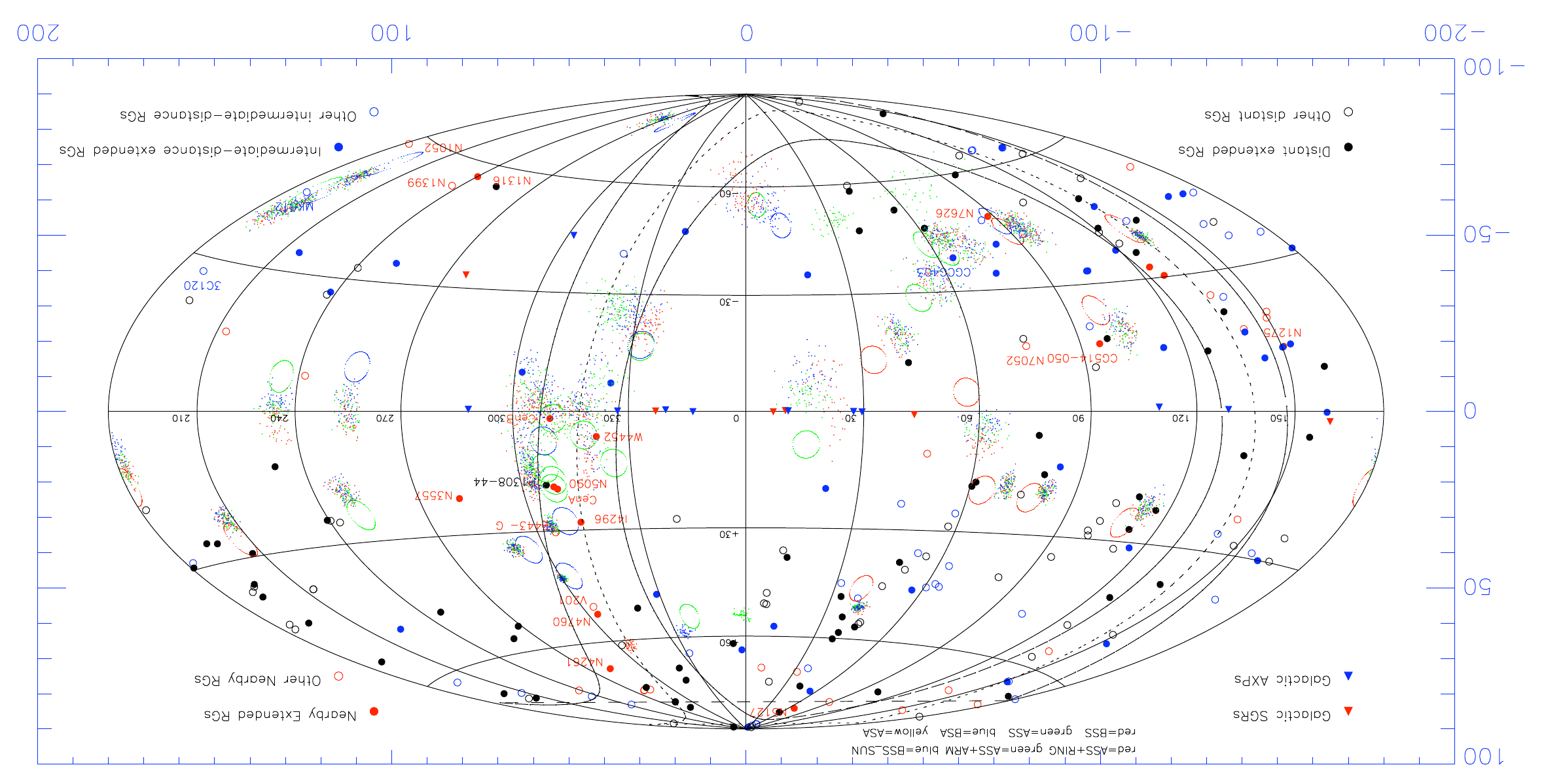}
}
\caption{Same as Fig.~\ref{figjet0123} for the three new models proposed
by S08: AS-S+RING (small red dots), AS-S+ARM (small green dots),
and BS-S (small blue dots). Here, as in S08, no ordered
dipole field component was used.
}
\label{figjet456}
\end{figure*}

\begin{figure*}[!htp]
\resizebox{4.8in}{!}{
\includegraphics[width=3.5in,angle=90,clip]{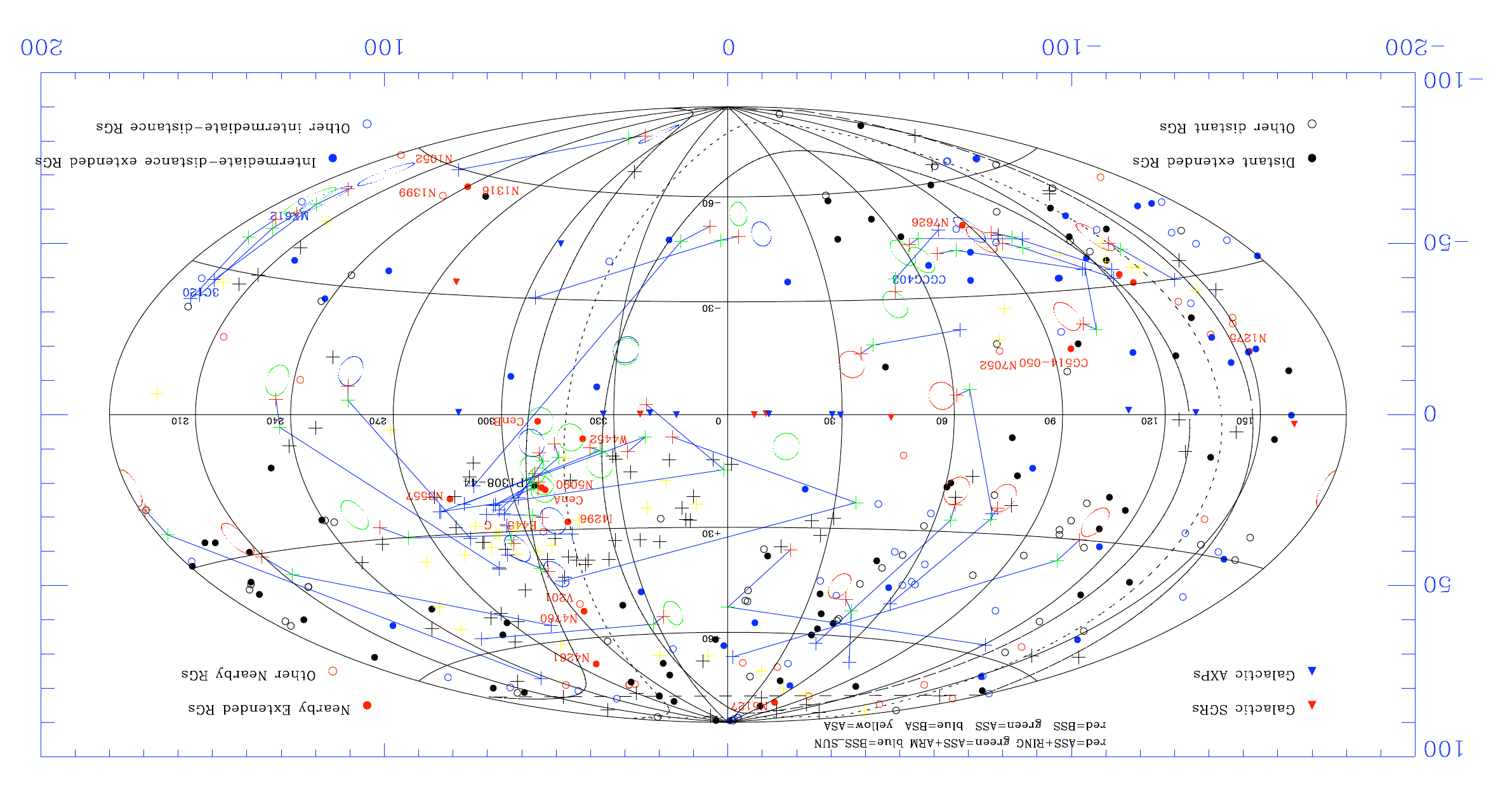}
}
\caption{A comparison of the measured earth-arrival directions of all UHECRs with
energies above 56~EeV detected by PAO and AGASA (blue, green, and red open circles
with radius 3.5\arcdeg\ as in Fig.\ref{figjet0123}) with the estimated outside-the-Galaxy 
arrival directions of the same UHECR (crosses) within the BS-S 
model for UHECR compositions of 
protons (red),
He (green),
Oxygen (blue),
Aluminum (yellow), and
Calcium and Iron (black). 
The first three compositions of the same UHECR are connected with blue lines while the 
heavier compositions are not connected to avoid overcrowding.
Other symbols and names in the figure are the same as in Fig. \ref{figjet0123}.
}
\label{figjetmulti}
\end{figure*}

\begin{figure*}[!ht]
\resizebox{4.8in}{!}{
\includegraphics[width=3.5in,angle=90,clip]{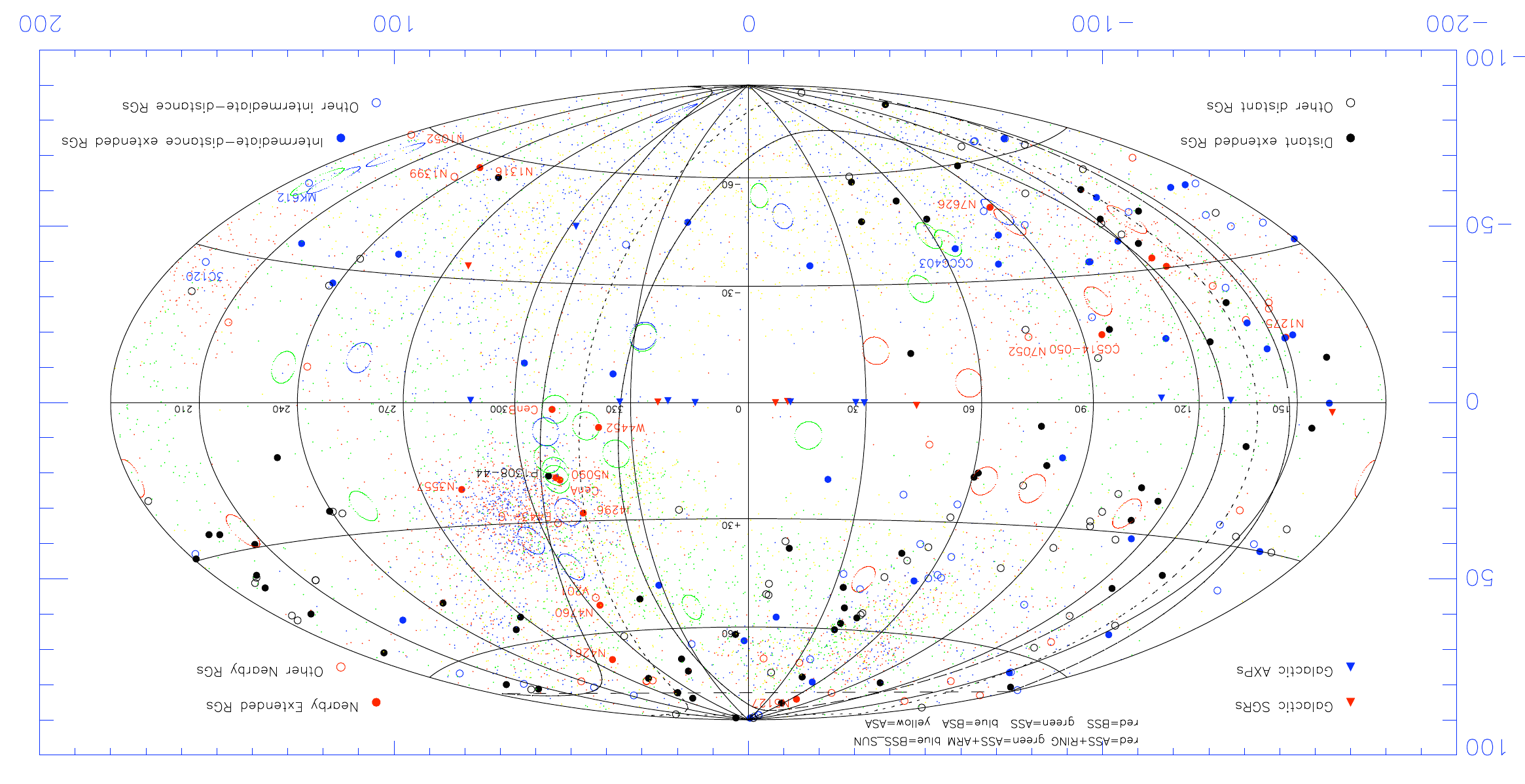}
}
\caption{Same as Fig.~\ref{figjet0123} but for a UHECR composition of Oxygen.
}
\label{figjet0123oxy}
\end{figure*}

\begin{figure*}[!ht]
\resizebox{4.8in}{!}{
\includegraphics[width=3.5in,angle=90,clip]{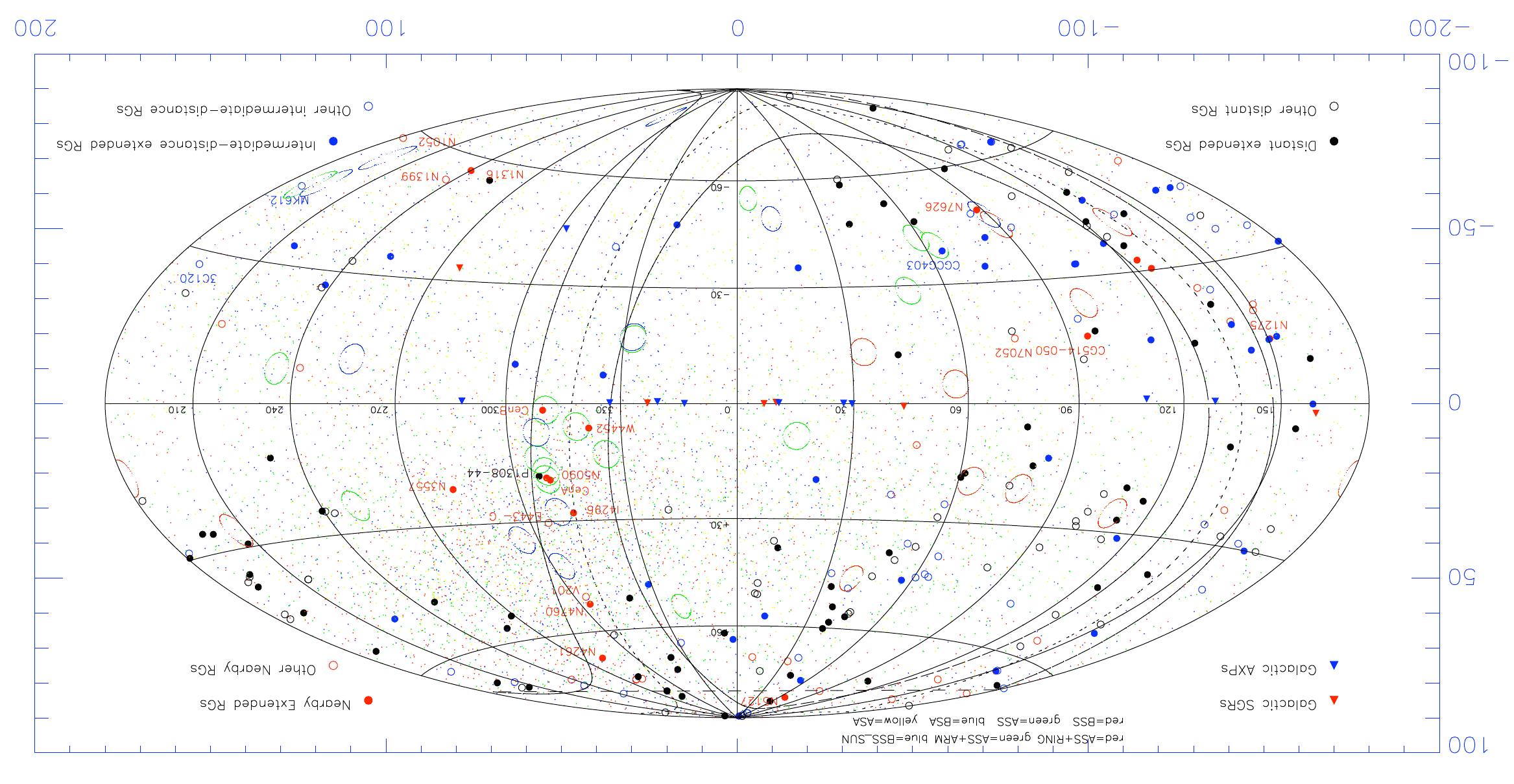}
}
\caption{Same as Fig.~\ref{figjet0123} but for a UHECR composition of Iron.
}
\label{figjet0123fe}
\end{figure*}

Graphical representations of our
simulation results are shown in Figures~\ref{figjet0123}
to \ref{figjet0123fe}. Coordinates of the arrival directions of the UHECRs detected by
PAO (blue for UHECRs with energy $\geq$ 75~EeV and green for UHECRs with energy
between 56 and 75~EeV) and AGASA (red) are shown as open circles of (arbitrary) radius 3.5\arcdeg.
The outside-the-Galaxy arrival directions of the UHECRs are plotted with 
small colored dots (50 per UHECR, each representing the result of one Monte Carlo simulation).
Figures~\ref{figjet0123} and \ref{figjetmulti} appear in the printed version of this manuscript while 
Figures~\ref{figjet456}, \ref{figjet0123oxy}, and  \ref{figjet0123fe} are available in the 
electronic version only.
The outside-the-Galaxy arrival directions were plotted in the order red, green, blue, and yellow. 
When one of these colors is not seen for a UHECR, it lies below one of the 
following colors in the sequence above. 
For example, the pairs BS-S (red), BS-A (blue) and AS-S (green), AS-A (yellow)
have the same deflection for positive values of Galactic latitude ($b$). 
To avoid overcrowding, results for the three models of S08  are 
shown separately from the other four models. 

Figures~\ref{figjet0123} to \ref{figjet0123fe} include the positions of
all galaxies with radio jets from the catalog of \citet{liuzha02}
within 500~kpc and the 16 confirmed and candidate SGRs and
AXPs in the Galaxy \citep{wootho06}.  
As in \citet{nagmat08} - but with different cutoffs - we 
have divided the galaxies with radio jets into three redshift bins: 
``nearby'' (D $\leq$ 75~Mpc; red circular symbols), 
``intermediate'' (75~Mpc $<$ D $\leq$ 200~Mpc; blue circular symbols),
and ``distant'' (200~Mpc $<$ D $\leq$ 500~Mpc; black circular symbols). 
In all redshift bins, we distinguish between galaxies with radio structures more 
extended than 180~kpc (``extended''; solid symbols) and 
those with radio structures less extended than 180~kpc (``compact''; open symbols).
Filled triangles are used for Galactic magnetars: red for SGRs
and blue for AXPs. Of the 16 magnetars, two are in the LMC and SMC and the others are
close to the plane of the Galaxy. 

\subsection{UHECR Trajectory Simulations: Protons}
\label{secresprotons}

Columns 5 to 7 of Tables~\ref{tabdef_auger} and \ref{tabdef_agasa}, present
the results of the deflection simulations for proton UHECRs within the BS-A GMF model.
Tabular results for other models are not shown for space reasons and can be 
requested from the authors.
The first four columns of the tables identify the UHECR by an index number, 
its arrival direction in Galactic coordinates, and its energy. 
Then, for each UHECR composition simulated, we list the average and r.m.s. 
of the deflection angles corresponding to the
50 outside-the-Galaxy arrival directions of each UHECR, followed by the
nearby extended radiogalaxy which best matches the fifty outside-the-Galaxy 
arrival directions (see the previous section for a definition of a match). 
A nearby extended radiogalaxy is only listed if more than four of the 50 outside-the-Galaxy
arrival directions are matched. 
For proton UHECRs, the average deflection and
its r.m.s. for each UHECR are both typically 1\arcdeg--5\arcdeg\ though values 
greater than 30\arcdeg\ are seen for a few UHECRs whose trajectories pass close 
to the Galactic plane and/or Galactic center.  

We emphasize four important results of the proton-UHECR trajectory simulations: 
(a)~even after considering errors in the model GMF and adding
    a relatively strong turbulent magnetic field component, the outside-the-Galaxy arrival 
    directions of the majority of proton-UHECRs are typically sufficiently concentrated 
    to allow meaningful comparisons with Galactic and extragalactic sources;
(b)~for a given proton-UHECR, the outside-the-Galaxy arrival directions are typically 
    significantly different for different GMF models. An assumption of the source
    population therefore allows constraints on competing models of the ordered GMF; 
(c)~almost all UHECRs with Galactic latitude $-25$\arcdeg$<b< 0$\arcdeg
    have trajectories which pass close to or cross
    the Galactic plane within certain GMF models, especially those which include an 
    ordered dipole field; 
(d)~as discussed in detail in Sec.~\ref{secresrg}, the match between
    proton-UHECRs and nearby extended radiogalaxies \citep{nagmat08} is maintained
    in most GMF models, and significantly strengthened in some GMF models. 

\subsection{UHECR Trajectory Simulations: Heavy Nuclei}
\label{secresheavy}

With increasingly heavy UHECR compositions, the increased deflections 
suffered by the UHECRs tend to scatter their out-of-Galaxy arrival
directions over larger regions of the sky. 
Figure~\ref{figjetmulti}, which compares the deflections
suffered by UHECRs in four models for different UHECR compositions,
emphasizes the importance of UHECR composition for source
identification.
The simulation in this figure does not include a Monte Carlo approach: i.e.
we did not consider GMF uncertainties or turbulent magnetic fields. 
We remark on several interesting features in Figure~\ref{figjetmulti}, especially
for compositions near Oxygen:
(a)~the three PAO UHECRs near $l=200$\arcdeg, $b=-45$\arcdeg\ move close to the
    radiogalaxy 3C120 for a UHECR composition of Oxygen;
(b)~the twin mystery PAO UHECRs near $l=335$\arcdeg, $b=-18$\arcdeg\ move toward the Galactic
    plane in the case of light compositions, and then toward the Cen~A concentration
    of radiogalaxies for an Oxygen composition. 
(c)~several UHECRs move toward the extended radiogalaxies NGC~315 and NGC~383
    (near position $l=130$\arcdeg, $b=-30$\arcdeg) for compositions near Oxygen.

A summary of the results of our Monte Carlo simulations for
heavy-nuclei UHECRs within the BS-A GMF model can be 
found in Tables~\ref{tabdef_auger} and \ref{tabdef_agasa}. Graphical
representations for Oxygen and Iron UHECRs 
within the BS-S, BS-A, AS-S, and AS-A models 
are presented in Figures~\ref{figjet0123oxy} and \ref{figjet0123fe}.
As expected, both the average and the r.m.s. of the deflection angles
corresponding to  the 50 outside-the-Galaxy
arrival directions for each UHECR increase with increasing UHECR mass 
(Tables~\ref{tabdef_auger} and \ref{tabdef_agasa}).
For compositions up to around Oxygen, the 50 outside-the-Galaxy arrival
directions corresponding
to each UHECR are typically still identifiably together on the sky - allowing
potential source searches (Figure~\ref{figjet0123oxy}). 
For compositions close to and higher than Ca, 
the fifty outside-the-Galaxy arrival directions corresponding to each UHECR are typically  
scattered over most of the sky. However, even in the case of Iron UHECRs
(Figure~\ref{figjet0123fe})
a concentration can still be seen in the region of the SuperGalactic
plane near to and north of Cen~A. 
 
\subsection{Source Populations: Centaurus A}
\label{secrescena}
 
The concentration of UHECRs in the Cen~A region has drawn a lot of attention
since its first report by the PAO team.
Our Monte Carlo simulations can be used to test whether Cen~A is the dominant, 
or only, source of UHECRs. 
The results of these tests are shown in 
Table ~\ref{tabcena_auger} for UHECRs detected by PAO and 
Table ~\ref{tabcena_agasa} for UHECRs detected by AGASA.
In these tables, the first column for each composition reports the fraction of
the total (1350 for PAO and 550 for AGASA) outside-the-Galaxy arrival directions
which fall within three 20\arcdeg\ circles centered on the Cen~A nucleus and its
northern and southern radio lobes. The second
column for each
composition shows the number of UHECRs (of the total of 27 for PAO and 11 for
AGASA), for which at least one of the fifty outside-the-Galaxy arrival directions
falls in the 20\arcdeg\ circles described above.

Within our simulation parameters, it is possible but unlikely that  Cen~A is the
source of all PAO-detected UHECRs: for proton UHECRs one third or less of the total 
outside-the-Galaxy arrival directions 
are within the 20\arcdeg\ circles around Cen~A in any GMF model, and almost two thirds
of PAO UHECRs do not have even one outside-the-Galaxy arrival direction within the above
area.  For heavier compositions a smaller fraction of total outside-the-Galaxy arrival directions
are within the 20\arcdeg\ circles around Cen~A but a larger fraction of PAO UHECRs, almost 
100\% for Ca and Fe, have at least one outside-the-Galaxy arrival direction which falls within the 
Cen~A circles. 
The latter result is due to the large scatter of the outside-the-Galaxy arrival directions of any
given heavy composition UHECR. In the case of AGASA UHECRs, the fraction of
outside-the-Galaxy arrival directions which fall within the Cen~A circles is very small for all compositions, 
though for heavy UHECR compositions, a large fraction or all UHECRs have at least one 
outside-the-Galaxy arrival direction falling within the Cen~A circles. 

If we consider all 300 outside-the-Galaxy arrival directions corresponding to each UHECR
(6 compositions, 50 outside-the-Galaxy arrival directions per composition) then within the models 
AS-S, AS-A, BS-S and BS-A, all PAO-detected UHECRs have at 
least one outside-the-Galaxy arrival direction which falls within the 20\arcdeg\ circles around Cen~A.
In the Sun et al. 2008 models a few (two to six) PAO UHECRs do not
satisfy the above criterion. For the AGASA-detected UHECRs, for the AS-S and AS-A
models all UHECRs have at least one outside-the-Galaxy arrival direction
which falls within the 20\arcdeg\ circles around  
Cen~A for some one composition, while in the other models between two and eight UHECRs
do not satisfy this criterion. 

\subsection{Source Populations: The Radio Galaxy Concentration around Cen~A}
\label{secresrgbox}

As \citet{nagmat08} have pointed out, the area around Cen~A hosts the highest
density of nearby extended radiogalaxies. We repeated the analysis of the previous
section, but using a box around all nearby extended radiogalaxies in this area.
For simplicity we used the following limits for the box:
Galactic longitude between 240\arcdeg\ and 360\arcdeg\, and 
Galactic latitude between $-10$\arcdeg\ and 70\arcdeg\ (the ``RG box'').
The results are listed in sub-columns 3 and 4 for each composition in 
Table ~\ref{tabcena_auger} for PAO UHECRs and 
Table ~\ref{tabcena_agasa} for UHECRs detected by AGASA.

As expected the match statistics are better than in the
case of only the 20\arcdeg\ circles around Cen~A and its radio lobes.  
The ``RG box'' contains between a quarter and a half of
all outside-the-Galaxy arrival directions for any one given composition. For light UHECRs at least half
of all PAO-detected UHECRs have at least one outside-the-Galaxy arrival direction which falls in
the RG box. For Ca and Fe compositions all UHECRs satisfy this criteria. 
In the case of AGASA UHECRs heavy compositions are required to provide a good
match between the UHECRs and the RG box. 

If we consider all 300 outside-the-Galaxy arrival directions corresponding to each UHECR
then within the models 
AS-S, AS-A, BS-S and BS-A, all PAO and AGASA UHECRs have at 
least one outside-the-Galaxy arrival direction which falls within the RG box. 
In the Sun et al. models a few (two to four) AGASA-detected UHECRs do not
satisfy the above criterion. 

\subsection{Source Populations: Nearby Extended Radio Galaxies}
\label{secresrg}

\citet{nagmat08} have previously argued that the arrival directions of
a subset of PAO-detected UHECRs are correlated with the directions of nearby extended
radiogalaxies. A distance of less than 3.5\arcdeg\ was found between 
the radio structures of six extended radiogalaxies (of the 10 ``visible'' to 
the PAO) and 8 UHECRs detected by PAO. 
Considering the full sky, we have a total of 15 nearby (D$<75$~Mpc) 
galaxies with extended ($>$ 180~kpc) radio structures, with the new galaxies 
listed in Sec.~\ref{secdata}, to be compared to 38 UHECRs.
Adding the AGASA-detected events, adds two new matches to earth-arrival 
directions of UHECRs: an AGASA event with original reported energy
E$=120~$EeV matched to NGC~7626 (which is also matched to a PAO-detected UHECR) and 
a match between an AGASA event with original reported energy E$=68~$EeV with CGCG~514$-$050
(this event does not appear in the figures as its 30\% reduced energy is below our
UHECR cutoff of 56~EeV). The correlation 
between nearby extended radiogalaxies and UHECRs detected by PAO and AGASA therefore
remains highly statistically significant even before considering deflections by the GMF.  
We remark that
the Supergalactic plane in the Cen~A region passes close to the Galactic longitudes
where the deflection produced by an ordered GMF with azimuthal 
symmetry is minimal \citep[e.g.][]{taksat08}. It may be this fortuitous coincidence
which allowed us to find several UHECR events ``matched'' to nearby extended radio galaxies 
in \citet{nagmat08}.

Including deflections of UHECRs by the GMF maintains, and in some models increases,
the correlation between UHECRs and nearby extended radiogalaxies. 
Table~\ref{tabellips}, in its first four columns, lists the statistics of matches between 
UHECRs and our sample of 15 nearby extended radiogalaxies for all compositions 
and models simulated. The BS-S and BS-A models clearly result in the best
match between nearby extended radiogalaxies and UHECRs. For proton UHECRs
the matches are few: in part due to using a rather strict match criterion
of 2.5\arcdeg. At heavy compositions a large fraction of UHECRs are matched
to nearby extended radiogalaxies. 

For proton-UHECRs, of the seven models tested the BS-S and BS-A models maintain
the high correlation between 
UHECRs and radiogalaxies in the Cen~A region 
(Figs.~\ref{figjet0123} to \ref{figjet0123fe}). 
Accounting for deflections within the BS-S and BS-A models lead to
a higher concentration of UHECRs around
Cen~A (the UHECR detected toward Cen~B is now also deflected up to Cen~A) and
the radiogalaxy WKK~4432. Applying deflections from the models of S08 do not 
result in good matches between UHECRs and radiogalaxies in this region,
though adding a dipole field to these models significantly improves
the matches. In the case of the three matched events near
NGC~7626, CGCG~413$-$019 and CGCG~514$-$050 \citep[see ][]{nagmat08}, all of BS-S,
BS-A, and AS-S are acceptable for UHECR-radiogalaxy matches;
however, the S08 models - with
or without a dipole component - do not provide consistently good matches.

For many other UHECRs in Figures~\ref{figjet0123} to \ref{figjet0123fe} 
we do not see obvious matches between the earth-arrival direction of UHECRs 
and nearby or intermediate (red and blue symbols) 
galaxies with radio jets. However, accounting for deflections by the GMF suggests several
new potential matches between proton-UHECRs and extended radio jets. Some interesting 
potential matches between galaxies with distance
$< 200$~Mpc and jet structures larger than 180~kpc are worth specifically noting:
(a)~we see possible matches between NGC~1275 and two
AGASA-detected UHECRs; 
(b)~CGCG~403$-$019, apart from a direct match to a PAO-detected UHECR,
is potentially the source of three additional  AGASA-detected UHECRs; 
(c)~the extragalactic Gamma-ray source with the highest maximum
    flux at HESS, Mrk~421, matches to  an AGASA UHECR; 
(d)~the BL~Lacs H~2356$-$309 and 1ES~0347$-$121, both detected by HESS, can also
    be matched to AGASA events.
Note that some of the AGASA events mentioned above do not appear in the figures
as their 30\% reduced energies are below 56~EeV.

To explain the majority of UHECRs as originating in 
galaxies with extended radio jets requires one of the following:
(a)~a stronger, or more ordered, intra-cluster and intergalactic magnetic 
    field resulting in extragalactic deflections greater than the current estimate 
    \citep[e.g.][]{haret02, dolet04}
    of a few degrees for UHECRs;
(b)~a larger mean free path or lower energy loss than current estimates of
    the GZK effect, such that more distant radiogalaxies
    can contribute to the detected UHECR events; 
(c)~a large fraction of heavy composition UHECRs. The higher deflections
    of heavy UHECRs in Galactic and extragalactic magnetic fields result in differences
    of several tens of degrees between the out-of-Galaxy and earth arrival directions.
    Recent results on UHECR composition from the Auger collaboration make this 
    scenario likely (PA09a; PA09b).

Given the small sample size of known UHECRs it is difficult to evaluate 
how including UHECR deflections in the GMF effects the correlation between UHECRs
and nearby massive spiral galaxies \citep{ghiet08}; this is best 
left for later analysis with a larger number of UHECRs. 
Our comparisons of UHECR outside-the-Galaxy arrival directions with the DRAGN sample,
and with the several catalogs of Gamma-Ray sources also revealed individually interesting
coincidences. However, their statistical significance is difficult to evaluate and we
do not discuss them here.

\subsection{Source Populations: Nearby Extended Radio Galaxies vs. Nearby Elliptical Galaxies}
\label{secresrc3}

\begin{figure*}[!ht]
\resizebox{4.8in}{!}{
\includegraphics[width=3.5in,clip]{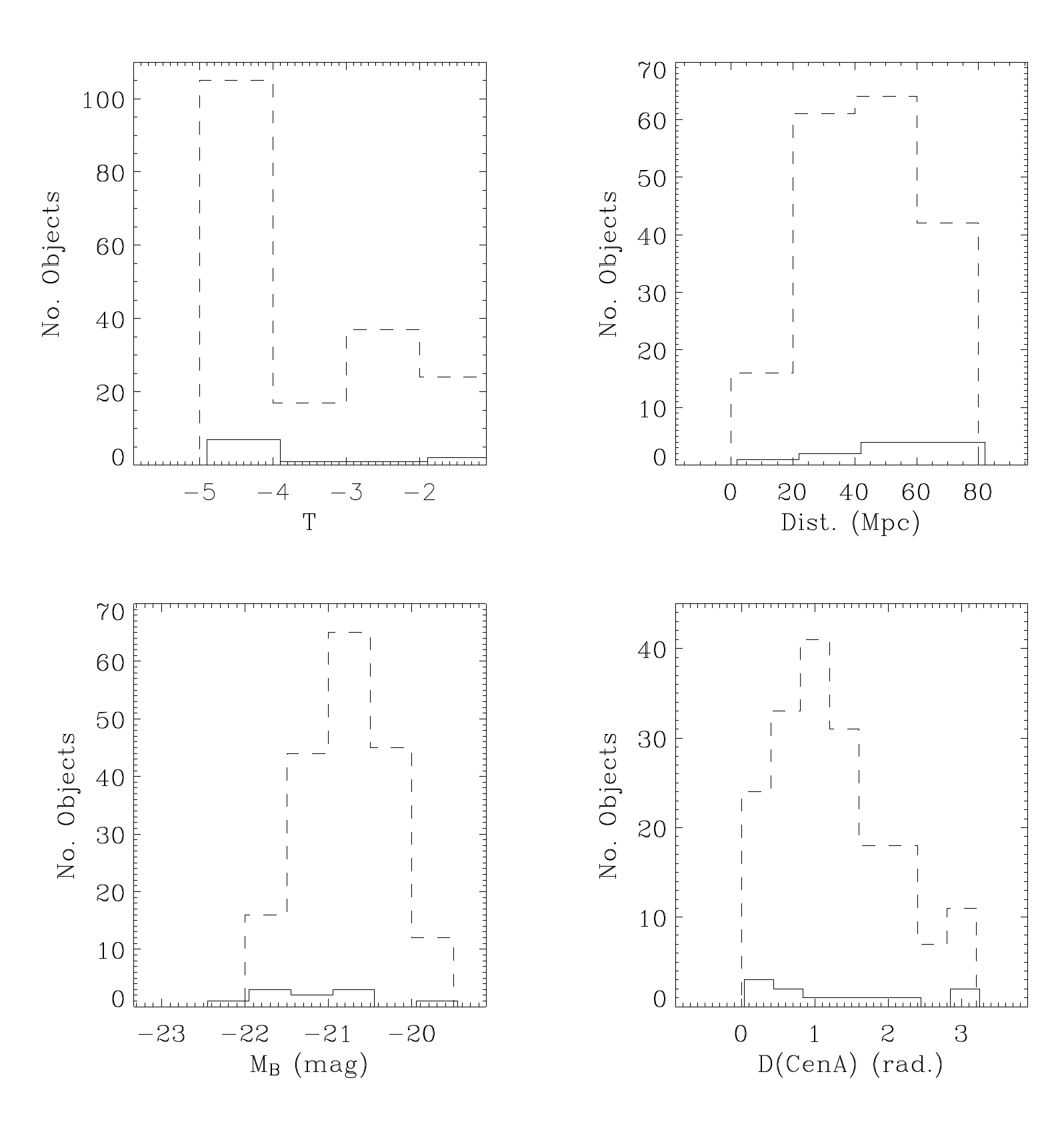}
}
\caption{Histograms of Morphological Type, Distance, Absolute B magnitude and
distance from Cen~A for the two samples of D $\leq$ 75~Mpc elliptical galaxies drawn 
from the RC3 catalog (see text): the 11 nearby RC3 ellipticals with extended radio structures
(solid line histograms) and the matched control sample of 183 nearby RC3 ellipticals without
extended radio structures.
}
\label{fighisto_compare}
\end{figure*}

Given the match between a subset of UHECRs and nearby extended radio galaxies,
it is relevant to test whether the potential association comes directly from
the presence of extended radio emission 
or from some underlying material associated with massive elliptical galaxies, for
e.g. luminous or dark matter concentrations. As a preliminary test, we have
used the RC3 catalog \citep{coret94} to extract all nearby (D$\leq$~75~Mpc) 
elliptical galaxies.
From this subset we identified all nearby RC3 ellipticals with extended radio jets 
and lobes, obtaining a list of 11 galaxies which 
includes all of our 15 nearby extended radio galaxies except NGC1275,
Cen~B, WKK~4552, and CGCG 514-050. 
The limits in RC3 parameters for this subset of ellipticals are:
distance D $\leq$ 75~Mpc, 
morphological type $-5 \leq$ T $\leq -2$, 
apparent B-band magnitude $7.69 \leq$ m$_{\rm B} \leq 13.38$, 
absolute B-band magnitude $-22.6 \leq$ m$_{\rm B} \leq -19.9$, 
and 
size  $1.2 \leq$ D$_{\rm 25}$ $\leq 2.42$, $0.02 \leq$ R$_{\rm 25}$ $\leq 0.2$,
      $1.21 \leq$ D$_{\rm 0}$ $\leq 2.45$. 
We then selected a control sample of all RC3 ellipticals without extended
radio jets and lobes which satisfied the above limits in RC3 measured parameters.
This resulted in a control sample of 183 ``radio-compact'' ellipticals.
Histograms comparing the distribution of distance, absolute magnitude,
size, and angular separation from Cen~A, of the two samples are shown in 
Fig.~\ref{fighisto_compare}. 
The two samples are reasonably well matched with a small tendency for the 
radio-extended ellipticals to have slightly higher absolute magnitudes and
to be slightly closer to Cen~A on average.

A comparison of the statistics of matches of the two elliptical samples to the 
outside-the-Galaxy arrival directions of PAO UHECRs is 
shown in Table~\ref{tabellips}.
When normalized for the number of galaxies in each sample, the matches
between UHECRs and extended radiogalaxies is significantly higher than
in the case of radio-compact ellipticals. This difference is most extreme
in the BS-S and BS-A models. 

\subsection{Source Populations: Potential Galactic Plane Sources}
\label{secresgal}

Very few UHECRs have been detected by PAO and AGASA in the region within 10\arcdeg\ 
of the Galactic plane. Accounting for deflections in the GMF, however, results in 
several ``Galactic plane crossing UHECRs'' in most GMF models. 
In particular, the BS-S and AS-S models, and the models of S08 
with or without a dipole component,
serve to move the arrival directions of 
several UHECRs at $-30<b<0$ towards the Galactic plane.

We searched for potential matches between the UHECR trajectories within the Galaxy and 
objects from the catalogs of Galactic microquasars and Galactic magnetars.
For the GMF models used, two or three UHECRs have trajectories which
pass close (within 1~kpc) to a SGR or AXP: in the case of SGR~1900+14, a good match is obtained
in all models except the S08 models without a dipole component.
In the case of SGR~1627-41, AS-S and AS-A model deflections lead to a connection
to the mysterious close pair of events detected by PAO. 
Several UHECR trajectories also pass close to confirmed or candidate Galactic
microquasars, and HESS-detected Galactic sources (e.g. Vela~X and Vela Junior).
Of course, fine tuning of the normalizations of the model permit closer matches;
given the many free parameters (not least the uncertain distances to these 
Galactic sources, and the mass of the UHECR).
We thus do not attempt a detailed study of the coincidences, but merely note 
that some UHECRs potentially cross the Galactic plane near 
candidate UHECR sources.

\section{Discussion \& Concluding Remarks}
\label{secdis}

The UHECR deflection simulations presented here, and the derived
outside-the-Galaxy arrival directions of UHECRs, highlight several
interesting points.
For light (proton or Helium) UHECRs, 
the relatively small dispersions in the outside-the-Galaxy arrival 
directions for a specific UHECR and GMF model confirm that tantalizing insights
into the source population of UHECRs can be gained with a small 
sample of UHECRs. While heavy composition UHECRs suffer deflections
of several tens of degrees the concentration of outside-the-Galaxy arrival directions 
in the quadrant containing the SuperGalactic plane near Cen~A also allows 
source population constraints.
Conversely, picking one or a few UHECR source populations permits an evaluation
of competing models of the ordered GMF.

For light compositions, the trajectories of several UHECRs pass close to or through 
the Galactic plane within several GMF models. 
Interestingly, several pass relatively close 
to SGRs, AXPs, and/or microquasars in the Galaxy. 
\citet{ghiet08} have argued that magnetars in
nearby galaxies could be responsible for the PAO detected UHECRs. The
integrated flux of HI emission from our galaxy 
is greater than the 
summed integrated HI fluxes of all massive spirals in the HIPASS survey considered
by \citet{ghiet08}. If UHECRs
originate in magnetars in steady state - rather than in formation - then it would not
be surprising to detect some magnetar-related UHECR events in the Galaxy.

In \citet{nagmat08} we claimed a correlation between nearby galaxies with extended 
radio jets and the earth-arrival direction of a subset of UHECRs. 
By fortuitous chance these matched pairs lie along the lines of Galactic 
longitude which suffer the minimum deflection by axisymmetric ordered Galactic magnetic
fields. The correlation is strengthened with the results presented here.
Accounting for deflections by BS-S and BS-A GMFs do not significantly change
the previously claimed matches in the case of proton UHECRs. 
Rather, using the BS-S model actually concentrates proton UHECRs closer to the nucleus of Cen~A, 
and also results in several new UHECR-radiogalaxy matches. 
The models of S08, with a dipole GMF component added, also maintain the previous matched pairs. 
Both the nearest radiogalaxy and
the nearest radio-extended BL~Lac are potentially sources of multiple UHECRs.
It is possible but unlikely that all UHECRs originate in Cen~A. 
In the best case, i.e. choosing the most convenient global GMF model for any given match,
about 30\% of UHECRs outside the range $-25$\arcdeg$<b<0$\arcdeg\ 
could be matched to a galaxy with extended radio jets or an extragalactic Gamma-ray
source in the case of proton only UHECRs. 
For heavy composition or varied composition UHECRs we cannot rule out the possibility
that all UHECRs originate in nearby galaxies with extended radio jets. We particularly
noted several interesting and new matches between UHECRs and radiogalaxies for
compositions near Oxygen.

A remaining question is whether the correlation between radiogalaxies and UHECRs
depends directly on the presence of extended radio jets and lobes or whether both
trace an underlying source population.
We have briefly addressed this issue in a comparative test of nearby ellipticals
with and without extended radio structure: there is a strong indication that, within
the BS-S and BS-A models of the GMF, the correlation with nearby extended
radiogalaxies is directly related to the radio jets rather than to an underlying
source population traced by massive elliptical galaxies.

If radiogalaxies, or sources traced by radiogalaxies,
are responsible for a subset of UHECRs, a BS-S or BS-A model for the GMF
is supported towards the Cen~A region, i.e. the northern hemisphere of the fourth
quadrant. The BS-S model (but not the BS-A) also allows a match between some UHECRs
and Galactic plane sources. The models of S08 are consistent with both of the
above only after we added a dipole field component (see Sect.\ref{secmonte} for why this is
dipole addition is not necessarily a self-consistent step).

Finally, we remark that this work identifies the subset of UHECRs
which can be matched to extended radiogalaxies using a light composition
(H or He) for UHECRs, and those which require a medium to heavy composition
(e.g. Table \ref{tabdef_auger}). 
The relative ratio of proton to heavy composition UHECRs then roughly agree
with the composition mix suggested by PA09a and PA09b.
It would be interesting to verify if the
X$_{\rm max}$ distributions of these two UHECR sub-samples confirm this composition 
difference.

\begin{acknowledgements}
We acknowledge funding from ALMA 3016013, ALMA 3107015, ALMA 3108022, 
Fondecyt 1080324, BASAL PFB-06/2007, and the FONDAP Center for Astrophysics.
This research has made use of the NASA/IPAC Extragalactic Database 
(NED) which is operated by the Jet Propulsion Laboratory, California 
Institute of Technology, under contract with the National Aeronautics 
and Space Administration.
\end{acknowledgements}

\clearpage
\onecolumn

\begin{table}
 \dummytable\label{tabdef_auger}
\end{table}
\begin{table}
 \dummytable\label{tabdef_agasa}
\end{table}
\begin{table}
 \dummytable\label{tabcena_auger}
\end{table}

\begin{table}
 \dummytable\label{tabcena_agasa}
\end{table}
\begin{table}
 \dummytable\label{tabellips}
\end{table}

\begin{figure}
\includegraphics[width=4.3in,clip]{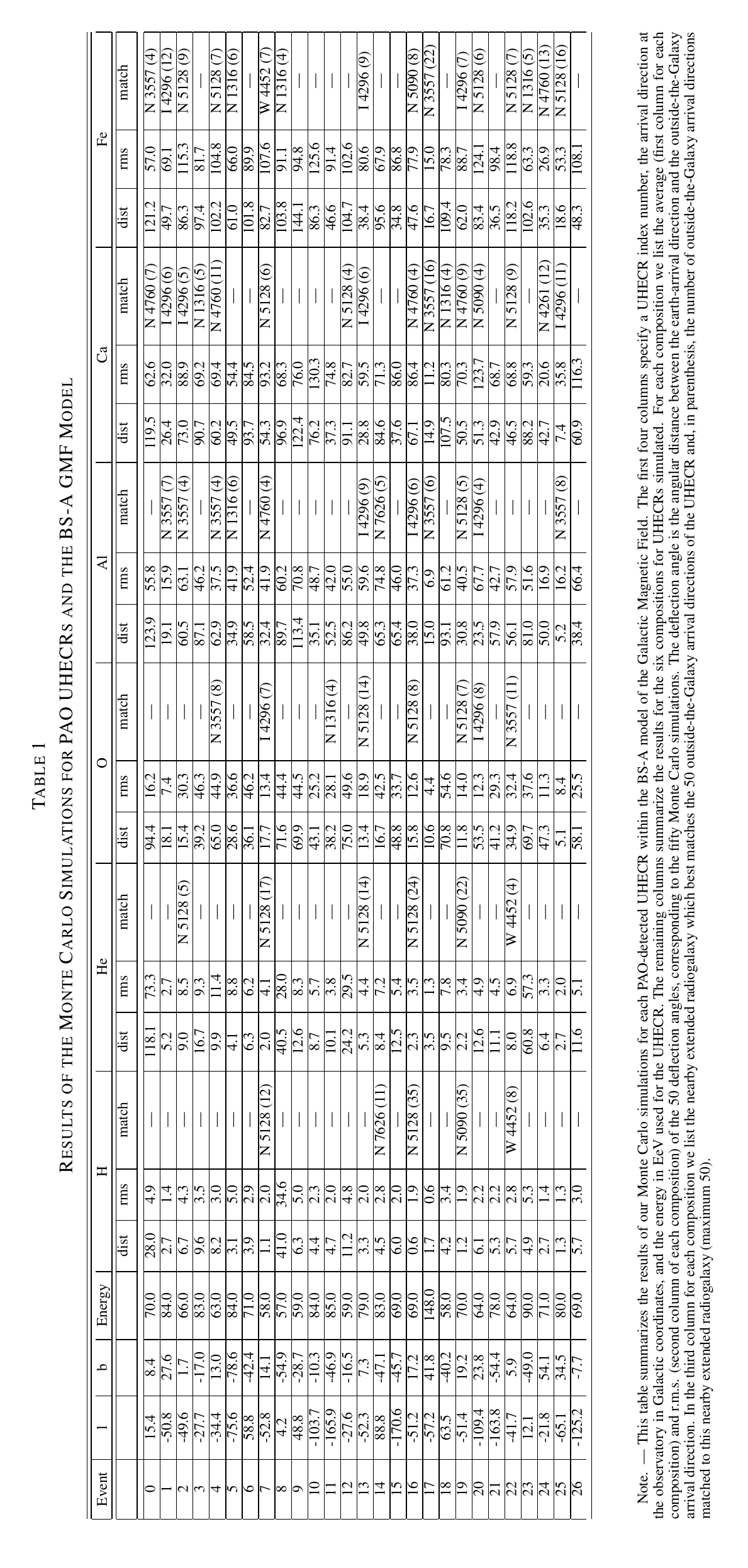}
\end{figure}

\begin{figure}
\includegraphics[width=2.5in,clip]{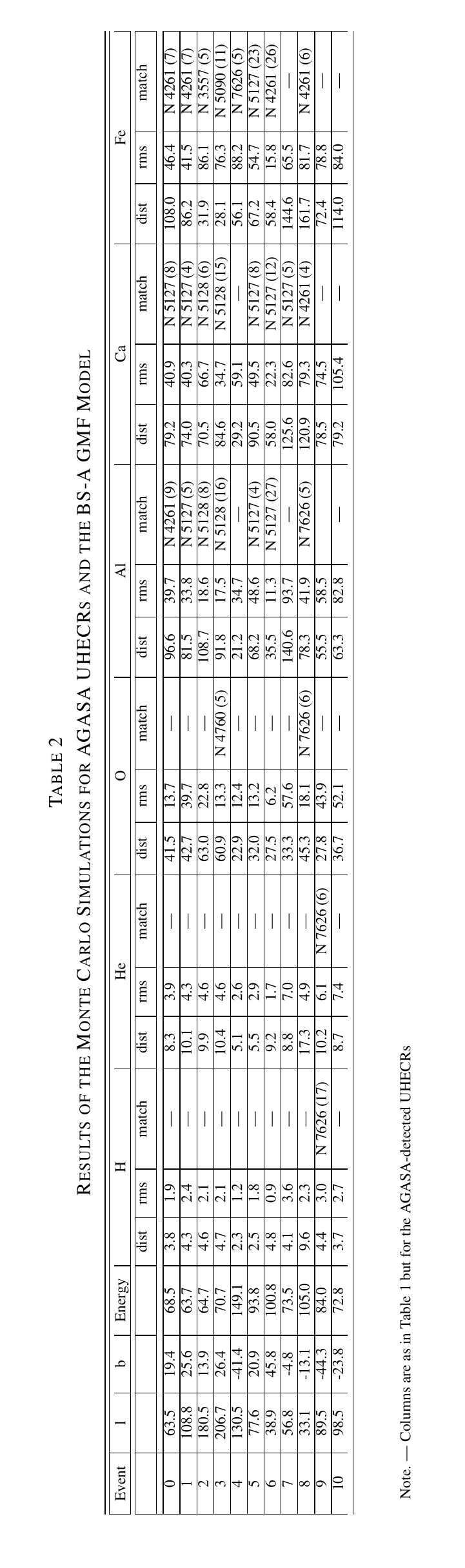}
\end{figure}

\begin{figure}
\includegraphics[width=3in,clip]{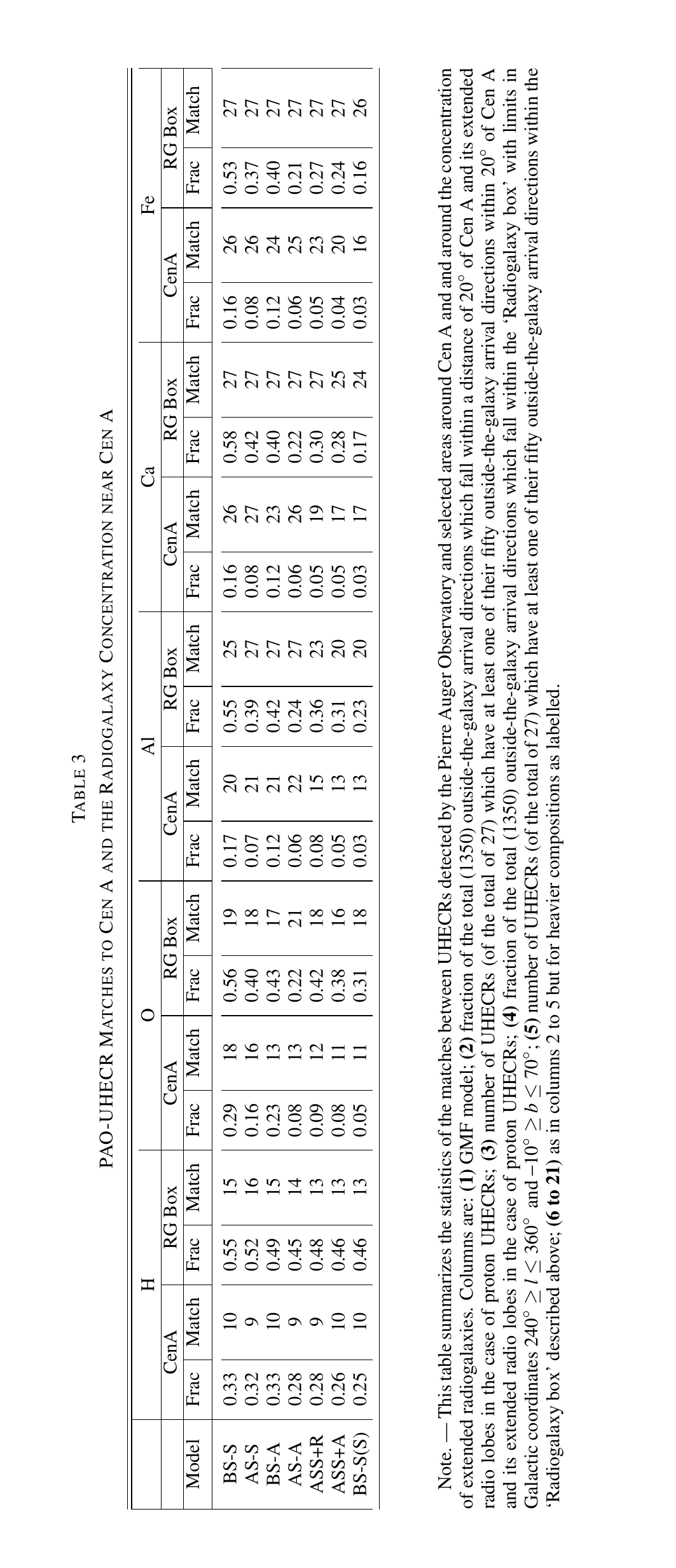}
\end{figure}

\begin{figure}
\includegraphics[width=2.5in,clip]{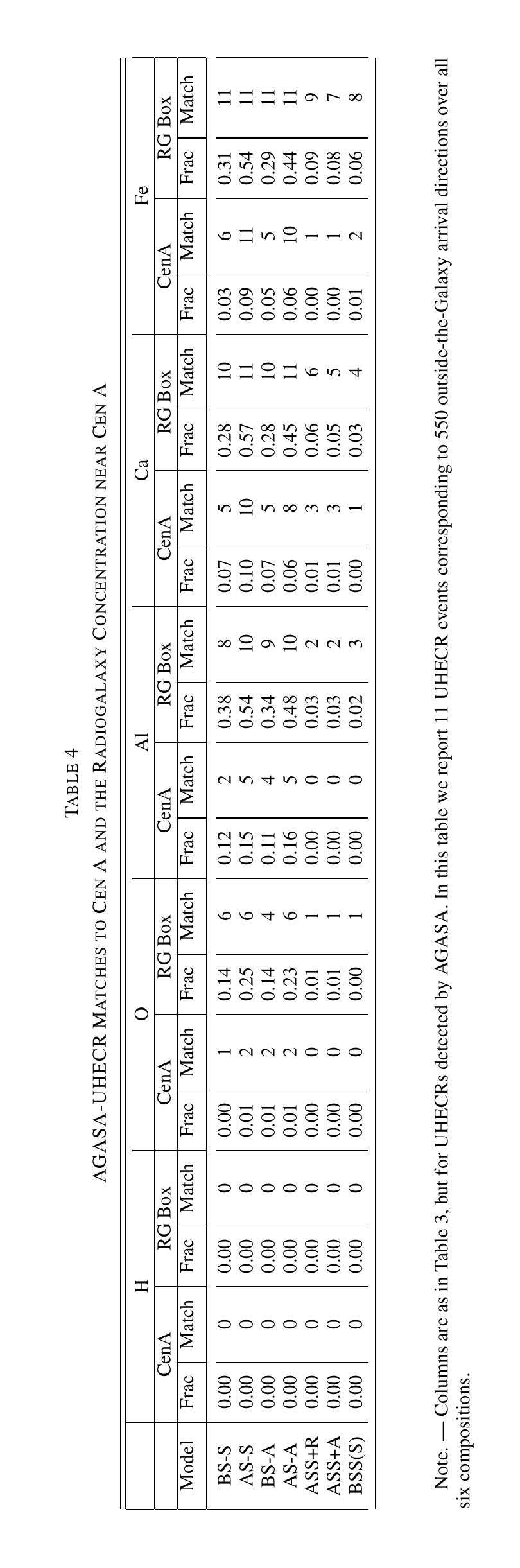}
\end{figure}

\begin{figure}
\includegraphics[width=4in,clip]{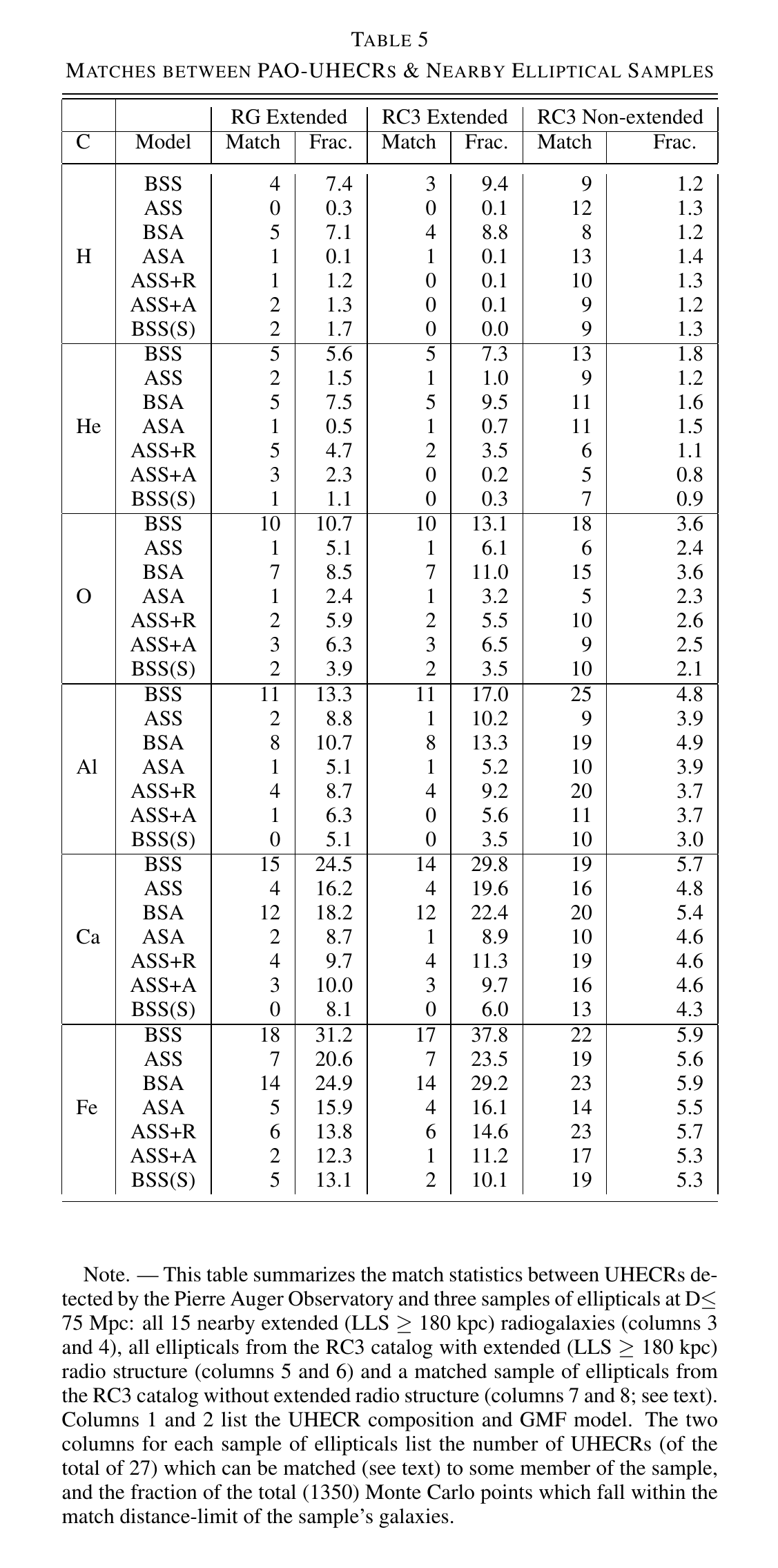}
\end{figure}

\end{document}